\documentclass[epj]{webofc}
\usepackage[utf8]{inputenc}
\usepackage[varg]{txfonts}   
\usepackage{booktabs}
\usepackage{xcolor}

\DeclareMathOperator{\Tr}{Tr}

\newcommand{\lb}{\langle \kern-.17em \langle} 
\newcommand{\rb}{\rangle \kern-.17em \rangle }

\newcommand{\beq}{\begin{eqnarray}}
\newcommand{\eeq}{\end{eqnarray}}
\newcommand{\nn}{\nonumber}

\newcommand\bs\boldsymbol

\definecolor{darkred}{rgb}{0.4,0.0,0.0}
\definecolor{darkgreen}{rgb}{0.0,0.4,0.0}
\definecolor{darkblue}{rgb}{0.0,0.0,0.4}
\usepackage[bookmarks,linktocpage,colorlinks,
    linkcolor = darkred,
    urlcolor  = darkblue,
    citecolor = darkgreen]{hyperref}
\wocname{EPJ Web of Conferences}
\woctitle{Lattice2017}
\begin{document}
%
\selectlanguage{english}
\title{%
Higgs compositeness in Sp(2N) gauge theories -- \\
Determining the low-energy constants with lattice calculations \thanks{Combined
  contributions of B. Lucini (\email{b.lucini@swansea.ac.uk}) and
  J.-W. Lee (\email{wlee823@pusan.ac.kr}).}
}
\author{%
\firstname{Ed} \lastname{Bennett}\inst{1}\fnsep\thanks{Funded by the Supercomputing Wales project, which is part-funded by the European Regional Development Fund (ERDF) via Welsh Government.} \and
\firstname{Deog Ki} \lastname{Hong}\inst{2}\fnsep\thanks{Supported in part by Korea Research Fellowship program funded by the Ministry of Science, ICT and Future Planning through the National Research Foundation of Korea (2016H1D3A1909283) and under the framework of international cooperation program (NRF-2016K2A9A1A01952069).} \and
\firstname{Jong-Wan}  \lastname{Lee}\inst{2,3}\fnsep\footnotemark[3] \and
\firstname{C.-J.~David} \lastname{Lin}\inst{4}\fnsep\thanks{Supported by Taiwanese MoST grant 105-2628-M-009-003-MY4.} \and \\
\firstname{Biagio} \lastname{Lucini}\inst{5}\fnsep\thanks{Supported in part by the Royal Society and the Wolfson Foundation.}\fnsep\thanks{Supported in part by the STFC Consolidated Grants ST/L000369/1 and ST/P00055X/1.} \and
\firstname{Maurizio} \lastname{Piai}\inst{6}\fnsep\footnotemark[6] \and
\firstname{Davide} \lastname{Vadacchino}\inst{1}\fnsep\footnotemark[2]\fnsep\footnotemark[6]
}

\institute{%
Swansea Academy of Advanced Computing, Swansea University, Singleton Park, Swansea, SA2 8PP, UK
\and
Department of Physics, Pusan National University, Busan 46241, Korea
\and
Extreme Physics Institute, Pusan National University, Busan 46241, Korea
\and
Institute of Physics, National Chiao-Tung University, Hsinchu 30010, Taiwan
\and
Department of Mathematics, Swansea University, Singleton Park, Swansea, SA2 8PP, UK
\and
Department of Physics, Swansea University, Singleton Park, Swansea, SA2 8PP, UK
}
\abstract{%
As a first step towards a quantitative understanding of the
SU(4)/Sp(4) composite Higgs model through lattice calculations, 
we discuss the low energy effective field theory resulting from the
SU(4) $\to$ Sp(4) global symmetry breaking pattern. We then consider
an Sp(4) gauge theory with two Dirac fermion flavours in the fundamental
representation on a lattice, which provides a concrete example of the 
microscopic realisation of the SU(4)/Sp(4) composite Higgs
model. For this system, we outline a programme of numerical simulations aiming at
the determination of the low-energy constants of the effective field
theory and we test the method on the quenched theory. We also report early
results from dynamical simulations, focussing on the phase structure
of the lattice theory and a calculation of the lowest-lying meson
spectrum at coarse lattice spacing. 
}
\maketitle
\section{Introduction}\label{intro}
In its conventional (and currently accepted) formulation, the Standard
Model includes an elementary Higgs field, whose condensate breaks the
electroweak symmetry. While this theory predicts all currently
measured observables to an impressive degree of accuracy, the
presence of an elementary scalar raises conceptual issues, one of the
main ones being the {\em hierarchy problem}. The latter expression
refers to the fact that the
mass of the Higgs boson (measured at around 125 GeV) is much smaller
than the natural cut-off scale, the Planck scale (which is of the of
$10^{19}$ GeV). A possible explanation of this puzzle could rely on
the assumption that the Higgs boson is not an elementary particle, but
a bound state resulting from a novel strong interaction. The current
experimental results show that no new particle is observed in a range
of energies up to a few TeV. Hence, even assuming that we can solve the hierarchy
problem with a non-elementary Higgs, we still need to face a question related to the naturaleness of the scales
involved, since the Higgs mass would anyhow be significantly lower
than that of other states of the novel strong interaction. This is
known as the {\em little hierarchy problem}.  

Higgs compositeness~\cite{Kaplan:1983sm}, i.e. the framework in which the Higgs boson
arises as a pseudo-Nambu-Goldstone boson (PNGB) of a novel strong
interaction, provides an elegant way to circumvent the hierarchy
problems of the Standard Model (see~\cite{Panico:2015jxa} for a recent
review). In order to study this approach to
compositeness, the first step is to understand the global symmetry
breaking pattern in the novel strong force, which has to be compatible
with the symmetries of the Standard Model. We start by considering the
patterns of possible global symmetry breaking in gauge theories with
gauge group ${\cal G}$ and $N_f$ flavours of Dirac fermions in the fundamental
representation. In this scenario, depending on G, relevant possibilities for
the symmetry breaking pattern G $\mapsto$ H are the following
(see e.g.~\cite{Peskin:1980gc}):
\begin{enumerate}
\item SU($N_f$)$_V$ $\times$ SU($N_f$)$_A$ $\mapsto$  SU($N_f$)$_V$ if
  ${\cal G}$ $\equiv$ SU($N$), $N > 2$; 
\item SU(2 $N_f$) $\mapsto$  SO(2 $N_f$) if the gauge group ${\cal G}$ is
  real; 
\item SU(2 $N_f$) $\mapsto$  Sp(2 $N_f$) if the gauge group ${\cal G}$ is
  pseudoreal. 
\end{enumerate}
In the PNGB scenario for the Higgs field, the electroweak gauge group of the
Standard Model is embedded in H, leading to the chain of embeddings
SU(2)$_L$ $\times$ U(1)$_Y$ $\subset$ H $\subset$ G. Hence, in this framework it is a necessary
condition that H be large enough to allow for embeddings of the Standard Model  SU(2)$_L$ $\times$ U(1)$_Y$. The four
components of the Standard Model Higgs fields are then identified with
four of the pions resulting from the global symmetry
breaking. Gauge theories whose dynamics is expected to be compatible
with the PNGB Higgs scenario have been discussed analytically
in~\cite{Barnard:2013zea,Ferretti:2013kya,Ferretti:2016upr,Vecchi:2015fma}. 

The Lattice provides a robust framework for a non-perturbative
analysis of the global symmetry breaking mechanism that
suggests the possible interpretation of the Higgs as a PNGB. Lattice
studies in this direction have started to appear only
quite recently. Examples of recent and current numerical work on
relevant models (some of which have been reviewed in~\cite{Svetitsky:2017xqk})
include~\cite{DeGrand:2015lna,Arthur:2016dir,DeGrand:2016htl,DelDebbio:2017ini,Ayyar:2017uqh,Ayyar:2017qdf}.   
In this contribution we provide the first (and still preliminary) numerical study of the Sp(4) gauge theory with $N_f = 2$ fundamental
Dirac fermions. Since Sp(4) is pseudoreal, according to our previous discussion, the expected global
symmetry breaking pattern in this case is
\beq
\mbox{SU(4)} \mapsto \mbox{Sp(4)} \ .
\eeq
As it will be clear from our discussion below, at this level this
symmetry breaking pattern has the requested properties: there are five broken
generators (hence, supporting via the Goldstone theorem the
existence of five massless bosons, four of which can be identified
with the Higgs doublet field) and the residual symmetry group Sp(4) has rank two,
which means that two independent SU(2) subgroups can be embedded into it,
hence allowing for the possibility of embedding an SU(2)$_L$ $\times$
U(1)$_Y$ group. 

The rest of this work is organised as follows. In Sect.~\ref{sec-1} we
describe the elementary gauge theory. For
the latter, using the framework of hidden local symmetry, an effective
Lagrangian is obtained in
Sect.~\ref{sec-2}. Section~\ref{lattice_action} is devoted to the
formulation of the lattice theory and of the observables we shall
compute. Our numerical results for the quenched setup and the
resulting low-energy constants in the effective field theory (EFT) approach
are discussed in Sect.~\ref{quenched}. Some exploratory numerical
calculations in the theory with dynamical fermions are reported
in Sect.~\ref{dynamical}. In Sect.~\ref{summary} we summarise our findings
and we outline future directions of our work. Two companion
contributions describe respectively the study of the
spectrum of Sp(4) Yang-Mills theory~\cite{davide} and some more technical
details (in particular, scale setting and topology) of Monte Carlo
calculations in Yang-Mills Sp(4) and in the theory with dynamical fermions~\cite{ed}.

\section{The model}\label{sec-1}
Before analysing the structure of the global symmetry breaking
pattern in Sp($2N$) gauge theories with $N_f$ fundamental Dirac fermions (of which
the specific model we investigate numerically, Sp(4) with two Dirac
fermions, is a particular case), we review the properties of Sp($2N$)
groups. 

Sp($2N$) can be defined as the subgroup of SU($2N$) whose
elements $U$ fulfil the condition
\beq
\label{eq:sp4:defining}
U \Omega U^{T} = \Omega  \ , \qquad 
\Omega = 
\left(
\begin{array}{cc}
0 & \mathbb{I}\\
- \mathbb{I} & 0
\end{array}
\right) \ , 
\eeq  
with $\mathbb{I}$ the $N \times N$ identity. $\Omega$ is known as the
{\em symplectic matrix}. The request of the invariance of $\Omega$
under the action of $U$ constrains the structure of the latter as follows:
\beq
U = 
\left(
\begin{array}{cc}
A & B\\
-B^{\ast}  & A^{\ast}
\end{array}
\right)
\ , \qquad \mathrm{with~~} A A^{\dag} + B B^{\dag} = \mathbb{I}
\mathrm{~~and~~} A^T B = B^T A \ , 
\eeq
where $A$ and $B$ are $N \times N$ matrices. For an element {\bf u} in
the algebra of Sp($2N$), this implies
\beq
{\bf u} = 
\left(
\begin{array}{cc}
{\bf a} & {\bf b}\\
{\bf b}^{\ast}  & - {\bf a}^{\ast}
\end{array}
\right)
\ , \qquad \mathrm{with~~} {\bf a} = {\bf a}^{\dag} 
\mathrm{~~and~~} {\bf b} = {\bf b}^T \ , 
\eeq
and again ${\bf a}$ and ${\bf b}$ are $N \times N$ matrices. Other properties of the
Sp($2N$) group that are of interest to us here are:
\begin{enumerate}
\item The dimension of the group is $N(2N+1)$;
\item The group is pseudoreal;
\item The group has rank $N$.
\end{enumerate}
In particular, the third property means that Sp($2N$) contains $N$
independent SU(2) subgroups.

As a way of an example, and as a reference for the numerical part, a
possible explicit embedding of the Sp(4) generators onto SU(4) is:
\beq
T^6&=&\frac{1}{2\sqrt{2}}
\left(
\begin{array}{cccc}
 0 & 0 & -i & 0 \\
 0 & 0 & 0 & -i \\
 i & 0 & 0 & 0 \\
 0 & i & 0 & 0
\end{array}
\right)
\,,~~~~~
T^7\,=\,\frac{1}{2\sqrt{2}}
\left(
\begin{array}{cccc}
 0 & 0 & 0 & -i \\
 0 & 0 & -i & 0 \\
 0 & i & 0 & 0 \\
 i & 0 & 0 & 0
\end{array}
\right)
\,, \\
\nonumber
T^8&=&\frac{1}{2\sqrt{2}}
\left(
\begin{array}{cccc}
 0 & -i & 0 & 0 \\
 i & 0 & 0 & 0 \\
 0 & 0 & 0 & -i \\
 0 & 0 & i & 0
\end{array}
\right)
\,,~~~~~
T^9\,=\,\frac{1}{2\sqrt{2}}
\left(
\begin{array}{cccc}
 0 & 0 & -i & 0 \\
 0 & 0 & 0 & i \\
 i & 0 & 0 & 0 \\
 0 & -i & 0 & 0
\end{array}
\right)
\,,
\\ \nonumber
T^{10}&=&\frac{1}{2}
\left(
\begin{array}{cccc}
 0 & 0 & 1 & 0 \\
 0 & 0 & 0 & 0 \\
 1 & 0 & 0 & 0 \\
 0 & 0 & 0 & 0
\end{array}
\right)
\,,~~~~~~~~~~~~~~
T^{11}\,=\,\frac{1}{2\sqrt{2}}
\left(
\begin{array}{cccc}
 0 & 0 & 0 & 1 \\
 0 & 0 & 1 & 0 \\
 0 & 1 & 0 & 0 \\
 1 & 0 & 0 & 0
\end{array}
\right)
\,, \\
\nonumber
T^{12}&=&\frac{1}{2}
\left(
\begin{array}{cccc}
 0 & 0 & 0 & 0 \\
 0 & 0 & 0 & 1 \\
 0 & 0 & 0 & 0 \\
 0 & 1 & 0 & 0
\end{array}
\right)
\,,~~~~~~~~~~~~~~
 T^{13}=\frac{1}{2\sqrt{2}}
\left(
\begin{array}{cccc}
 0 & 1 & 0 & 0 \\
 1 & 0 & 0 & 0 \\
 0 & 0 & 0 & -1 \\
 0 & 0 & -1 & 0
\end{array}
\right)
\,,
\\ \nonumber
T^{14}&=&\frac{1}{2\sqrt{2}}
\left(
\begin{array}{cccc}
 1 & 0 & 0 & 0 \\
 0 & -1 & 0 & 0 \\
 0 & 0 & -1 & 0 \\
 0 & 0 & 0 & 1
\end{array}
\right)
\,,~~~~~
T^{15}\,=\,\frac{1}{2\sqrt{2}}
\left(
\begin{array}{cccc}
 1 & 0 & 0 & 0 \\
 0 & 1 & 0 & 0 \\
 0 & 0 & -1 & 0 \\
 0 & 0 & 0 & -1
\end{array}
\right)
\,.
\eeq

A choice for the generators in the coset is
\beq
T^1&=&\frac{1}{2\sqrt{2}}
\left(
\begin{array}{cccc}
 0 & 1 & 0 & 0 \\
 1 & 0 & 0 & 0 \\
 0 & 0 & 0 & 1 \\
 0 & 0 & 1 & 0
\end{array}
\right)
\,,~~~~~
T^2\,=\,\frac{1}{2\sqrt{2}}
\left(
\begin{array}{cccc}
 0 & -i & 0 & 0 \\
 i & 0 & 0 & 0 \\
 0 & 0 & 0 & i \\
 0 & 0 & -i & 0
\end{array}
\right)
\,, \\ \nonumber
T^3&=&\frac{1}{2\sqrt{2}}
\left(
\begin{array}{cccc}
 1 & 0 & 0 & 0 \\
 0 & -1 & 0 & 0 \\
 0 & 0 & 1 & 0 \\
 0 & 0 & 0 & -1
\end{array}
\right)
\,,~~~~~
T^4\,=\,\frac{1}{2\sqrt{2}}
\left(
\begin{array}{cccc}
 0 & 0 & 0 & -i \\
 0 & 0 & i & 0 \\
 0 & -i & 0 & 0 \\
 i & 0 & 0 & 0
\end{array}
\right)
\,, \\ \nonumber
T^5&=&\frac{1}{2\sqrt{2}}
\left(
\begin{array}{cccc}
 0 & 0 & 0 & 1 \\
 0 & 0 & -1 & 0 \\
 0 & -1 & 0 & 0 \\
 1 & 0 & 0 & 0
\end{array}
\right) \ .
\eeq
For the algebra of SU(4) and Sp(4) we have respectively
\begin{itemize} 
\item $su(4) = \mathrm{Span}\{T^1, T^2, \dots, T^{15}\}$ \ ,
\item $sp(4) = \mathrm{Span}\{T^6, T^7, \dots, T^{15}\}$ \ .
\end{itemize}
~\\
The properties
\beq
\nonumber
&& \Omega\, T^A + T^{AT} \Omega = 0\qquad A\,=\,6, \ \dots,\ 15 \ , \\ 
\nonumber
&& \Omega \,T^A - T^{AT} \Omega = 0\qquad A\,=\,1, \ \dots,\ 5
\eeq
easily follow from Eq.~(\ref{eq:sp4:defining}). 

\begin{table}
\caption{The field content of the theory. The two columns indicate the
  tranformation laws under Sp(4) gauge and global SU(4).}
\centering
\label{tab:1}
\begin{tabular}{|c|c|c|c|}
\hline\hline
{\rm ~~~Fields~~~} &Sp(4)  &  SU(4)\cr
\hline
$V_{\mu}$ & $10$ & $1$ \cr
$Q$ & $4$ & $4$ \cr
\hline
$\Sigma_0$ & $1$ & $6$\cr
$M$ & $1$ & $6$\cr
\hline\hline
\end{tabular}
\end{table}

The elementary theory we have investigated contains two flavours of Dirac fermions $Q^i_a$
(with the index $i$ identifying the flavour and $a$ the colour Sp(4)
component) transforming under the fundamental representation of Sp(4)
and the gauge fields $V_{\mu}$. Due to the fact that Sp(4) is pseudoreal, the global
symmetry is enhanced from SU(2)$_L$  $\times$ SU(2)$_R$ to SU(4). 
With the above conventions on the generators
understood and the transformation laws of the fields under Sp(4) gauge
and global SU(4) given in Tab.~\ref{tab:1}, the Lagrangian density of the
elementary theory is given by 
\beq
{\cal L}=i\,\overline{Q^i}_{\,a}\,\gamma^{\mu}\,(D_{\mu}Q^i)^a\,-\,m\,\overline{Q^i}_{\,a}Q^{i\,a}\,-\,\frac{1}{2}\Tr V_{\mu\nu} V^{\mu\nu}\,.
\eeq
$V_{\mu \nu}$ is the field strength of the Sp(4) gauge bosons.

We can trade the two Dirac spinors for four left-handed Weyl
components $q^{i a}$ (see e.g.~\cite{Lee:2017uvl}). We then define the
antisymmetric colour combination of two quarks ({\em diquark}) 
\beq
\Sigma_0^{\,\,nm} \equiv\Omega_{ab} q^{n\,a\,T} \tilde{C} q^{m\,b}
\ ,
\eeq
where the colour indices $a$ and $b$ are contracted, while the flavour
indices $n$ and $m$ (both running from one to four) are free.

When using the diquark fields as variables, the mass term is
conveniently rewritten as $M\equiv m\, \Omega$. Hence, if a condensate
forms, $\Sigma_0 \propto M$, which implies the global symmetry
breaking pattern SU(4) $\mapsto$ Sp(4). 

\section{The effective field theory}\label{sec-2}
The Chiral Lagrangian corresponding to our global symmetry breaking
pattern can be written down using standard techniques. First, we
define the field $\Sigma$ transforming under the
antisymmetric representation of SU(4), i.e.
\beq
\Sigma \rightarrow  U \Sigma U^T  \ , 
\eeq
with $U$ an element of global SU(4). If $\Sigma$ acquires a
non-null vacuum expectation value ({\em vev}), we will have (modulo an
SU(4) rotation) 
\beq
\langle \Sigma \rangle \propto \Omega \ , 
\eeq
i.e. this vev will be responsible for the breaking SU(4) $\mapsto$
Sp(4). We can now introduce the parameterisation
\beq
\Sigma =e^{\frac{i\pi}{f}}\Omega
e^{\frac{i\pi^T}{f}}\,=\,e^{\frac{2i\pi}{f}}\Omega\,=\,\Omega
\,e^{\frac{2i\pi^T}{f}} \ .
\eeq
In the chiral EFT approach, at the lowest order we can write
\beq
{\cal L}_0&=&\frac{f^2}{4}\Tr\left\{\frac{}{}\partial_{\mu}\Sigma
  \,(\partial^{\mu}\Sigma)^{\dagger}\right\}\\
\nonumber
&=&\Tr
\left\{\partial_{\mu}\pi\partial^{\mu}\pi\right\}\,+\,\frac{1}{3f^2}\Tr\left\{\frac{}{}\left[\partial_{\mu}\pi\,,\,\pi\right]\left[\partial^{\mu}\pi\,,\,\pi\right]\right\}\,+\,\cdots
\ , 
\eeq
from which we immediately get the identification  $f = f_\pi$. 

In order to introduce a mass term, we define the spurion field $M$ transforming
as $M\rightarrow U^{\ast} M U^{\dagger}$, in terms of which we get the
massive part of the Lagrangian
\beq
{\cal L}_m&=&-\frac{v^3}{4}\Tr \left\{M\,\Sigma\right\}\,+\,{\rm h.c.}
= 2m v^3 \,-\,\frac{m v^3}{f^2}\Tr \pi^2\,+\,\frac{m\,v^3}{3f^4}\Tr
(\pi\pi\pi\pi)+\cdots \ .
\eeq
From this equation we deduce that the pions are degenerate and that the GMOR relation
\beq
m_{\pi}^2f_{\pi}^2={m \,v^3}
\eeq
holds.

\begin{figure}[tb] 
  \centering
  \includegraphics[scale=1.5]{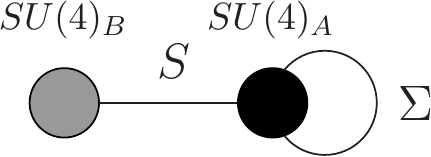}
  \caption{The moose diagram representing the hidden local symmetry
    construction for our model.}
  \label{fig-1}
\end{figure}

The framework of hidden local symmetry~\cite{Bando:1984ej} provides a prescription
to include in the effective Lagrangian the vector ($\rho$) and and the
axial-vector ($a_1$) mesons. The procedure (represented by the moose diagram of Fig.~\ref{fig-1})
consists in the following steps:
\begin{enumerate}
\item We redefine the SU(4) group under which $\Sigma$ transforms in the
  antisymmetric representation as SU(4)$_A$: 
\beq
\Sigma\,\rightarrow\, U_A \Sigma U_A^T ;
\eeq
\item We introduce the scalar mesons $\sigma$ by defining $S =
  e^{\frac{2i\sigma}{F}}$ and couple $S$ bilinearly to the
  (conjugated) SU(4)$_A$ on the right and to an additional global
  SU(4)$_B$ on the left:
\beq
S&\rightarrow & U_B\, S \,U_A^{\dagger}\, ;
\eeq
\item We gauge SU(4)$_A$ and a SU(2)$_L $ $\times$ U(1)$_Y$ $\subset$
  Sp(4) $\subset$ SU(4)$_B$.
\end{enumerate}

The covariant derivatives of the gauge theory take the forms
\beq
D_{\mu}S&=&\partial_{\mu} S+i  \left(g W_{\mu} + g^{\prime}B_{\mu}
\right)S-iSg_{\rho} A_{\mu} \ ; \\
D_{\mu}\Sigma&=&
\partial_{\mu} \Sigma + i \left[\frac{}{}(g_{\rho}A_{\mu})
  \Sigma\,+\,\Sigma (g_{\rho}A_{\mu})^T\right] \ .
\eeq
It is now straightforward to determine the field content of the
theory, consisting of $19$ gauge bosons (15 $A_{\mu}$, three $W_{\mu}$ and
one $B_{\mu}$) and $20$ pseudoscalar fields (five $\pi$ and 15 $\sigma$)

Following a standard procedure that we will detail elsewhere, we
arrive at the following effective Lagrangian:
\beq
{\cal L}&=&-\frac{1}{2}\Tr\left.B_{\mu\nu}B^{\mu\nu}\right.-\frac{1}{2}\Tr\left.W_{\mu\nu}W^{\mu\nu}\right.-\frac{1}{2}\Tr\left.A_{\mu\nu}A^{\mu\nu}
-\frac{\kappa}{2}\Tr\left[ A_{\mu\nu} \Sigma (A^{\mu\nu})^T \Sigma^{\ast}\right]\right.\nonumber\\
&&+\frac{f^2}{4}\Tr\left[\frac{}{}D_{\mu}\Sigma\,(D^{\mu}\Sigma)^{\dagger}\right]\,+\,\frac{F^2}{4}\Tr\left[\frac{}{}D_{\mu}S\,(D^{\mu}S)^{\dagger}\right]\nonumber\\
&&+b \frac{f^2}{4}\Tr\left\{D_{\mu}(S\Sigma )\left(D^{\mu}(S\Sigma)\right)^{\dagger}\right\}\,+\,
c\frac{f^2}{4} \Tr\left\{D_{\mu}(S\Sigma S^T)\left(D^{\mu}(S\Sigma S^T)\right)^{\dagger}\right\}\,\nonumber\\
&&-\frac{v^3}{8}\Tr\left\{\frac{}{}M\, S\, \Sigma\,S^{T}
\right\}\,+\,{\rm h.c.} \label{Eq:L} \nonumber \\
&&- \frac{v_1}{4} \Tr\left\{\frac{}{} M\, (D_{\mu} S)\, \Sigma \, (D^{\mu} S)^T \right\}\,
 -\frac{v_2}{4} \Tr\left\{\frac{}{} M\, S\,(D_{\mu}  \Sigma) \, (D^{\mu} S)^T \right\}\,+\,{\rm h.c.}\,\nonumber\\
 &&
 -\frac{y_3}{8}\Tr\left\{A_{\mu\nu}\Sigma\left[(A^{\mu\nu})^TS^T M S-S^T M S A^{\mu\nu}\right]\right\}\,+\,{\rm h.c.}\nonumber\\
  &&
 -\frac{y_4}{8}\Tr\left\{A_{\mu\nu}\Sigma\left[(A^{\mu\nu})^TS^T M S+S^T M S A^{\mu\nu}\right]\right\}\,+\,{\rm h.c.}\,
\nonumber\\
&&
-\frac{v_5^2}{128}\left(\Tr M S \Sigma S^T \,+\, {\rm
    h.c.}\frac{}{}\right)^2 \ .
\eeq
In the low momentum expansion, the order is determined by pairs of
derivatives, which are equivalent to mass insertions. Using this
counting rule, we identify the leading terms in the
first four lines of the previous equation. The corresponding couplings
are the parameters $\kappa,
f, F, b, c, v$. The last four lines contain some of the subleading
terms. Computing all the subleading terms is a demanding
calculation. Hence, we have determined only those that we expect should be
relevant for the study of mesonic two-point functions. 

It is important to stress that the
effective theory introduced above is only valid if $g_{\rho}^2/(4\pi)^2
\ll 1$. This can be checked {\em a posteriori}, by fitting the
couplings and 
the other parameters of the low-energy theory through a match with
observables derived from the fundamental theory. The ability of the
derived low-energy theory to describe observables of the elementary
theory at large distances is also a test of validity of the performed truncation.
Writing down the relevant observables in the EFT
and performing the relevant calculations (e.g. determining the  positions of the
poles of the two-point functions in momentum space in order to identify meson masses),
we arrive to relations between parameters of the effective Lagrangian
and physical observables. These include the following:
\beq
M_{\rho}^2&=&\frac{1}{4(1+\kappa+m\,y_3)} {g_{\rho}}^2 \left(b
  f^2+F^2+2 m v_1\right)\, \ ,\nn \\
M_{a_1}^2&=&\frac{1}{4(1-\kappa-m \,y_4)} {g_{\rho}}^2 \left(b f^2+F^2+2 m v_1\right)+\frac{g_{\rho}^2}{1-\kappa} \left(f^2+m
   (v_2-v_1)\right)\, \ , \nn \\
   f_{\rho}^2&=&
   \frac{1}{2} \left(b f^2+F^2+2 m v_1\right)\, \ ,\nn \\
   f_{a_1}^2&=&
   \frac{\left(b f^2-F^2+2 m (v_1-v_2)\right)^2}{2 \left((b+4) f^2+F^2-2 m
   v_1+4 m v_2\right)}\, \ , \nn \\
   f_0^2&=&\frac{}{}F^2+(b+2c)f^2\, \ , \nn \\
   f_{\pi}^2&=&f_0^2-f_{\rho}^2-f_{a_1}^2\, \ .
\label{eq:observables_eft}
\eeq
Other calculable quantities that for brevity we do not write
explicitly here include $m_{\pi}$, $g_{\rho \pi \pi}$, which is 
related to $g_{\rho}$, the Weinberg sum rules and the $S$
parameter. Explicit expressions for those quantities will be reported
in a forthcoming publication.  

In the next two sections we describe lattice calculations that will be used
to match the EFT with the elementary theory.

\section{Lattice action and observables}
\label{lattice_action}

The lattice action considered in this work is the standard plaquette action along with 
the unimproved Wilson Dirac fermions in the fundamental representation,
\beq
S[U]=\beta \sum_{\mu<\nu} \left(1-
\frac{1}{4} \textrm{Re}\,\textrm{Tr} U_\mu(x)U_\nu(x+\hat{\mu})U^\dagger_\mu (x+\hat{\nu}) U^\dagger_\nu (x))
\right)
+a^4\sum_x \bar{\psi}(x) (D+m_0) \psi(x),
\label{eq:lattice_action}
\eeq
where $\beta=8/g^2$ and $m_0$ are the bare lattice gauge coupling and the bare fermion mass, respectively. 
Note that the link variables $U_\mu(x)$ are elements of the Sp(4) group. 
The massless Wilson-Dirac operator is given by 
\beq
a D \psi(x)=4 \psi(x)-\frac{1}{2}\sum_\mu
\{
(1-\gamma_\mu)U_\mu(x)\psi(x+\hat{\mu})+(1+\gamma_\mu)U(x-\hat{\mu})\psi(x-\hat{\mu})
\},
\eeq
where $a$ is the lattice spacing. 
The implementations of the heat bath (HB) algorithm for the pure gauge
model and of the hybrid Monte-Carlo (HMC) algorithm in 
the dynamical fermion case of the lattice system described by the action in
Eq.~(\ref{eq:lattice_action}) are discussed respectively in~\cite{davide} and in~\cite{ed}.  
An essential technical difference from the case in which the gauge
group is SU(N) is a consequence of the fact that Sp(4) configurations span the maximal symplectic
subgroup space of SU(4), which means that reunitarisation must be replaced by
resymplectisation.   

To take full advantage of the EFT results derived in Sect.~\ref{sec-2}, 
we measure the masses and decay constants of pseudoscalar,  vector, 
and axial-vector mesons. As such mesons can be defined in the isotriplet channel, 
it is sufficient to consider flavoured particles with 
the corresponding interpolating operators given by
\beq
\mathcal{O}_\textrm{PS}(x)=\bar{Q}^i(x)\gamma_5 Q^j(x),~~
\mathcal{O}_\textrm{V}(x)=\bar{Q}^i(x)\gamma_\mu Q^j(x),~~
\mathcal{O}_\textrm{AV}(x)=\bar{Q}^i\gamma_5\gamma_\mu Q^j,
\eeq
where $i\neq j$ are flavour indices. 
Although not explicitly shown, summations over the colour and spinor indices are understood. 

For the given mesonic operator $\mathcal{O}_M$ the Euclidean two-point correlation function 
on a $T\times L^3$ lattice is defined as
\beq
C_{\mathcal{O}_M}(\vec{p},t)=\sum_{\vec{x}} e^{-i\vec{p}\cdot \vec{x}} 
\langle 0 | \mathcal{O}_M(\vec{x},t) \mathcal{O}_M^\dagger(\vec{0},0) | 0\rangle, 
\label{eq:corr}
\eeq
where the meson mass $m_M$ is extracted from the asymptotic behaviour 
of the correlator at large Euclidean time and zero momentum
\beq
C_{\mathcal{O}_M}(t)\xrightarrow{t\rightarrow \infty}\langle 0 |\mathcal{O}_M | M \rangle \langle 0 |\mathcal{O}_M | M \rangle^* 
\frac{1}{m_M L^3} \left[e^{-m_M t}+e^{-m_M(T-t)}\right].
\label{eq:meson_corr}
\eeq
To calculate the decay constants $f_M$ we parameterise the matrix elements of mesons 
as follows:
\beq
\langle 0 | \overline{Q_1} \gamma_5 \gamma_\mu Q_2 | PS \rangle = if_\pi p_\mu,~~
\langle 0 | \overline{Q_1} \gamma_\mu Q_2 | V \rangle = if_\rho m_\rho \epsilon_\mu,~~
\langle 0 | \overline{Q_1} \gamma_5 \gamma_\mu Q_2 | AV \rangle = if_{a_1} m_{a_1} \epsilon_\mu,
\label{eq:matrix_element}
\eeq
where $\epsilon_\mu$ is the polarisation vector 
transverse to the four-momentum, i.e. $\epsilon_\mu p^\mu=0$. 
The meson states $|M\rangle$ 
are the self-adjoint isospin fields, 
not the charged meson fields, 
as the corresponding pion decay constant in QCD is 
$f_\pi \simeq 93\,\textrm{MeV}$ in our convention. 
For the vector and axial-vector mesons, 
it is straightforward to calculate the masses and decay constants 
from the asymptotic forms of $C_{\mathcal{O}_M}(t)$ in Eq.~(\ref{eq:meson_corr})
using the parameterisations in Eq.~(\ref{eq:matrix_element}). 
In the case of the pseudoscalar meson, 
along with the pseudoscalar two-point correlation function, 
we consider an additional correlation function 
\beq
C_{\Pi}(\vec{p},t)=\sum_{\vec{x}} e^{-i\vec{p}\cdot \vec{x}} 
\langle 0 | [\overline{Q_1} \gamma_5 \gamma_\mu Q_2(\vec{x},t)]\,[\overline{Q_1} \gamma_5 Q_2(\vec{0},0)] | 0\rangle.
\label{eq:axial_corr_1}
\eeq
Using the fact that this correlation function is dominated by 
the lowest mass pseudoscalar state at large time, one can find
\beq
C_{\Pi}(\vec{p},t)\xrightarrow{t\rightarrow\infty}
\frac{if_\pi \langle 0 | \mathcal{O}_{PS} | PS \rangle^*}{L^3}\left[
e^{-m_\pi t}-e^{-m_\pi (T-t)}\right], 
\label{eq:axial_corr}
\eeq
where $m_\pi$ and $f_\pi$ are determined from the simultaneous fits 
for $C_\Pi(\vec{p},t)$ and $C_{\textrm{PS}}(\vec{p},t)$. 

To convert the measured quantities in lattice units to physical ones 
in the continuum, 
an appropriate scale setting and renormalisation are required. 
For the latter, in this work, we renormalise the decay constants 
by performing a perturbative one-loop matching. 
For Wilson fermions the decay constants receive finite multiplicative renormalisation 
as
\beq
f^{\rm ren}_\pi=Z_A f_\pi,~~~f^{\rm ren}_\rho = Z_V f_\rho,~~~{\rm and}~~~ f^{\rm ren}_{a_1}=Z_A f_{a_1},
\label{eq:one_loop_matching}
\eeq
where the matching coefficients are calculated in Ref.~\cite{Martinelli:1982mw}
\beq
Z_A&=&1+C(F)\left(\Delta_{\Sigma_1}+\Delta_{\gamma_5\gamma_\mu}\right)\frac{\tilde{g}^2}{16\pi^2}, \nn \\
Z_V&=&1+C(F)\left(\Delta_{\Sigma_1}+\Delta_{\gamma_\mu}\right)\frac{\tilde{g}^2}{16\pi^2},
\label{eq:zfactor}
\eeq
with $C(F)=5/4$ for the Sp(4) theory with fundamental fermions. 
The $\Delta$'s are results of relevant one-loop integrals performed
numerically. Their values are
\beq
\Delta_{\Sigma_1}=-12.82,~~~
\Delta_{\gamma_\mu}=-7.75,~~~
\Delta_{\gamma_5\gamma_\mu}=-3.00.
\eeq
Furthermore, as the perturbative expansion for Wilson fermions with a bare lattice coupling 
converges slowly, we use the coupling renormalised by a simple Tadpole improvement 
\beq
\tilde{g}=\frac{g^2}{\langle \textrm{tr} \mathcal{P} \rangle/4},
\eeq
where $\mathcal{P}$ is the usual plaquette operator. 

For the scale setting we adapt the gradient flow method with the choice of 
$\mathcal{W}(\bar{t})\equiv t\,\textrm{d}\mathcal{E}(\bar{t})/\textrm{d}t$, 
where as the scale-setting observable a symmetric four-plaquette clover has been used for the discretisation 
of the action density $\mathcal{E}(\bar{t})$ at the fictitious flow
time $\bar{t}$. A comparison of this with other potential choices is discussed
in~\cite{ed}.  

\section{Numerical results in the quenched limit}
\label{quenched}

Although the effects of fermions are not properly captured in the quenched approximation, 
the latter is often studied to understand the qualitative features of the full theory, 
an obvious advantage being the possibility of performing calculations
using comparatively smaller computational resources. 
For instance, in the quenched framework one can test the low-energy EFT in Sect.~\ref{sec-1} and demonstrate its use 
by calculating the associated mesonic observables. 
To this end, using the HB algorithm implemented to the HiRep code~\cite{DelDebbio:2008zf,DelDebbio:2009fd}, 
we generate two ensembles of pure Sp(4) gauge theory at $\beta=7.62$ and $8.0$ 
on a $48\times 24^3$ lattice with $200$ independent configurations in
each case. To convert the measured lattice quantities to the physical ones, 
we first calculate the gradient flow scale $w_0$ and the one-loop 
matching constants, for which the numerical results are found in Tab.~\ref{tab:zfactor}.
Note that our choice for the reference scale in the scale-setting procedure 
is $\mathcal{W}(\bar{t})|_{\bar{t}=w_0^2}=\mathcal{W}_0=0.35$. 

\begin{table}
\begin{center}
\begin{tabular}{c|c|c|c|c}
\hline\hline
 & $~~~w_0~~~$ &~~~plaquette~~~ & $~~~~~Z_V~~~~~$ & $~~~~~Z_A~~~~~$ \cr \hline
$b=7.62$ & $1.448(4)$ & $0.60190(19)$ & $0.71599(9)$ & $0.78157(7)$ \cr
$b=8.0$ & $2.308(6)$ & $0.63074(13)$ & $0.74185(5)$ & $0.80146(4)$ \cr
\hline\hline
\end{tabular}
\end{center}
\caption{
\label{tab:zfactor}
Gradient flow scales $w_0$, plaquette values and one-loop matching
factors $Z_V$ and $Z_A$ at the values of $\beta$ used in our quenched calculation.
}
\end{table}

At various fermion masses we calculate the two-point correlation functions 
of mesons in Eq.~(\ref{eq:corr}) at zero momentum using stochastic wall sources. 
If the Euclidean time is large enough, in principle, 
one can extract the masses and the decay constants using the asymptotic 
forms of the correlation functions in
Eqs.~(\ref{eq:meson_corr})~and~(\ref{eq:axial_corr}). If the time
separation is not long enough, however, one should take into account
the contribution of excited states, which translates into considering
a multi-exponential behaviour.
In most cases we find that single and two exponential fits are sufficient to 
describe the numerical data at large time. 
A correlated fit with $\chi^2$ minimisation is used to calculate the masses and 
decay constants, where the statistical uncertainties are estimated 
using the standard bootstrapping technique. 
We show our numerical results in
Tabs.~\ref{tab:quenched_b7.62}~and~\ref{tab:quenched_b8.0}. 
In order to apply an analysis based on the continuum EFT to the numerical results, 
we should express observables in physical units by multiplying by the (length) scale $w_0$ 
the quantities having mass dimension one and take into account the multiplicative 
renormalisation for the decay constants. 

\begin{table}
\caption{%
\label{tab:quenched_b7.62}%
Masses and bare decay constants of pseudoscalar, vector and axial-vector mesons 
for quenched calculations performed at $\beta=7.62$ on a $48\times 24^3$ lattice. 
}
\begin{center}
\begin{tabular}{ccccccc}
\hline \hline
~~$m_0$~~ & ~~ $m_{\textrm{PS}}^2$ ~~ & ~~ $m_{\textrm{V}}^2$ ~~ & ~~ $m_{\textrm{AV}}^2$ ~~ & ~~ $f_{\textrm{PS}}^2$ ~~ & ~~ $f_{\textrm{V}}^2$ ~~ & ~~ $f_{\textrm{AV}}^2$ ~~ \\ \hline
-0.65 & 0.4325(5) & 0.5087(8) & 1.04(4) & 0.01451(9) & 0.0376(3) & 0.021(4) \\
-0.7 & 0.3042(4) & 0.3916(11) & 0.943(29) & 0.01246(8) & 0.0365(4) & 0.029(3) \\
-0.73 & 0.2318(4) & 0.3272(14) & 0.862(20) & 0.01101(8) & 0.0354(5) & 0.0313(20) \\
-0.75 & 0.1856(4) & 0.2875(15) & 0.831(16) & 0.00995(8) & 0.0346(4) & 0.0352(16) \\
-0.77 & 0.1409(4) & 0.2485(19) & 0.769(22) & 0.00879(8) & 0.0329(6) & 0.0350(22) \\
-0.78 & 0.1191(4) & 0.2312(21) & 0.796(17) & 0.00822(8) & 0.0327(6) & 0.0415(16) \\
-0.79 & 0.0977(4) & 0.2115(26) & 0.772(20) & 0.00760(8) & 0.0314(8) & 0.0418(19) \\
-0.8 & 0.0765(4) & 0.193(3) & 0.748(24) & 0.00698(8) & 0.0301(10) & 0.0417(23) \\
-0.81 & 0.0553(4) & 0.175(5) & 0.73(3) & 0.00635(9) & 0.0285(14) & 0.042(3) \\
-0.815 & 0.0446(4) & 0.166(7) & 0.70(4) & 0.00606(9) & 0.0280(19) & 0.040(4) \\
-0.82 & 0.0328(4) & 0.158(13) & 0.70(6) & 0.00572(15) & 0.028(4) & 0.041(6) \\ \hline \hline
\end{tabular}
\end{center}
\end{table}

\begin{table}
\caption{%
\label{tab:quenched_b8.0}%
Masses and bare decay constants of pseudoscalar, vector, and axial-vector mesons
for quenched calculations performed at $\beta=8.0$ 
on a $48\times 24^3$ lattice. 
}
\begin{center}
\begin{tabular}{ccccccc}
\hline \hline
~~$m_0$~~ & ~~ $m_{\textrm{PS}}^2$ ~~ & ~~ $m_{\textrm{V}}^2$ ~~ & ~~ $m_{\textrm{AV}}^2$ ~~ & ~~ $f_{\textrm{PS}}^2$ ~~ & ~~ $f_{\textrm{V}}^2$ ~~ & ~~ $f_{\textrm{AV}}^2$ ~~ \\ \hline
-0.45 & 0.5556(12) & 0.5837(13) & 0.868(13) & 0.00832(14) & 0.01528(27) & 0.0057(5) \\
-0.5 & 0.4244(11) & 0.4551(14) & 0.734(11) & 0.00771(14) & 0.0149(3) & 0.0073(6) \\
-0.55 & 0.3037(8) & 0.3383(12) & 0.593(11) & 0.00691(10) & 0.0145(3) & 0.0084(7) \\
-0.6 & 0.1937(8) & 0.2325(14) & 0.465(12) & 0.00567(8) & 0.0131(3) & 0.0096(8) \\
-0.625 & 0.1437(7) & 0.1862(16) & 0.405(13) & 0.00494(8) & 0.0125(4) & 0.0102(10) \\
-0.64 & 0.1156(7) & 0.1612(15) & 0.363(20) & 0.00449(7) & 0.0124(3) & 0.0099(18) \\
-0.65 & 0.0974(7) & 0.1448(16) & 0.349(15) & 0.00414(7) & 0.0120(3) & 0.0109(13) \\
-0.66 & 0.0812(5) & 0.1302(14) & 0.343(14) & 0.00397(5) & 0.0123(3) & 0.0128(12) \\
-0.67 & 0.0642(5) & 0.1179(18) & 0.318(18) & 0.00352(4) & 0.0123(4) & 0.0123(17) \\
-0.68 & 0.0479(4) & 0.1040(22) & 0.323(13) & 0.00319(4) & 0.0122(5) & 0.0150(11) \\
-0.69 & 0.0318(4) & 0.0907(28) & 0.304(19) & 0.00270(5) & 0.0115(5) & 0.0151(17) \\ \hline \hline
\end{tabular}
\end{center}
\end{table}

\begin{figure}[t]
\begin{center}
\includegraphics[width=.49\textwidth]{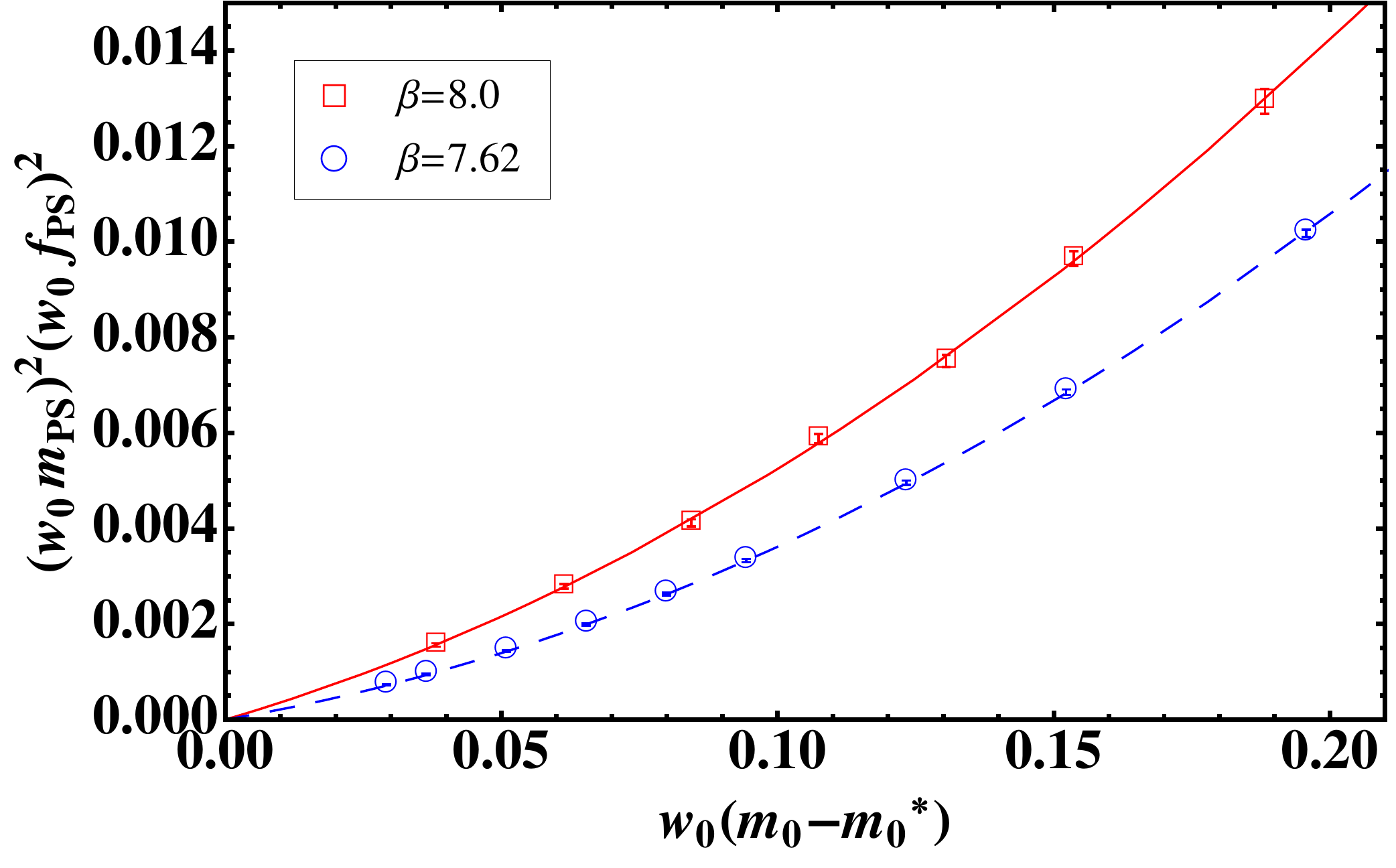}
\includegraphics[width=.49\textwidth]{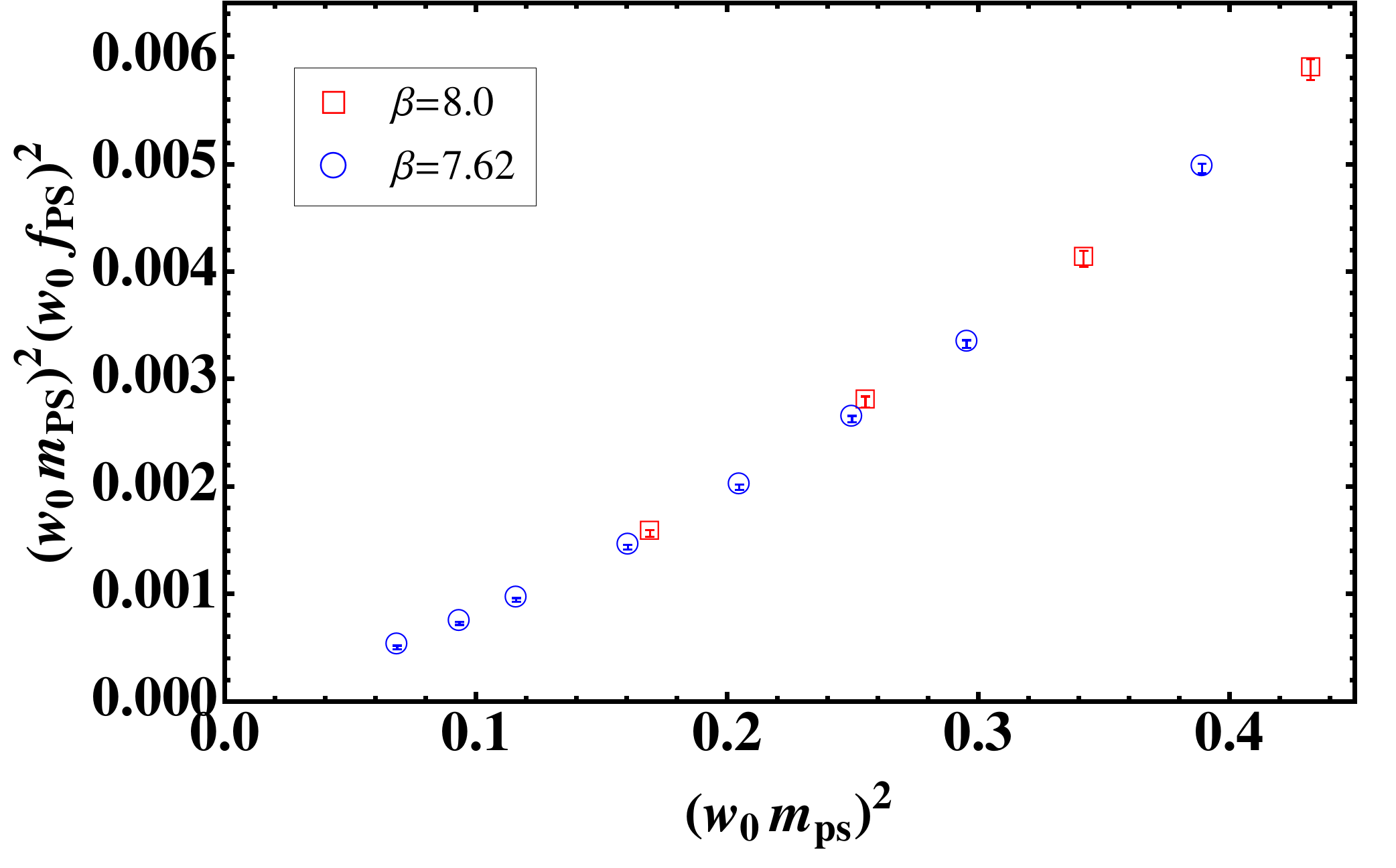}
\caption{%
Product of mass squared and decay constant squared for the pseudoscalar meson. 
Scale setting with the gradient flow method and 
perturbative one-loop matching for the decay constants are understood. 
In the left panel, $m_{\textrm{PS}}^2 f_{\textrm{PS}}^2$ is measured with respect to the bare fermion mass 
shifted by the critical mass, $m_0^*$, with the latter determined from quadratic fits. 
The solid and dashed lines are the fit results. 
In the right panel, the same combination is measured with respect to
the squared pseudoscalar mass. All quantities are expressed in
units of $w_0$.
}
\label{fig:GMOR}
\end{center}
\end{figure}

By restricting our attention to the pseudoscalar mesons and using the GMOR relation 
we first determine the critical fermion mass at which $m_{\textrm{PS}}$ vanishes. 
As the numerical results of $m_{\textrm{PS}}^2 f_{\textrm{PS}}^2$ do not
show the validity of the leading order GMOR relation, 
we perform quadratic fits over the bare mass range $-0.82\leq m_0 \leq -0.73$ for $\beta=7.62$ and $-0.69\leq m_0 \leq -0.625$ 
for $\beta = 8.0$, 
and find $-w_0\,m_0^*=-1.214(22)$ and $-w_0\,m_0^*=-1.636(27)$, respectively. 
The resulting GMOR relation is illustrated in the left panel of Fig.~\ref{fig:GMOR}. 
Note that the results with different lattice coupling (or equivalently different lattice spacing) 
cannot be compared directly, as the quark mass renormalisation depends
on $\beta$. One can overcome this problem by simply considering 
a physical (and hence free from renormalisation) quantity such as the pseudoscalar mass $m_{\textrm{PS}}$. 
In the right panel of Fig.~\ref{fig:GMOR} we replot $m_{\textrm{PS}}^2 f_{\textrm{PS}}^2$ with respect to $m_{\textrm{PS}}$, 
and find that the GMOR relations at the two lattice couplings are statistically identical. 
In other words, the lattice artefacts due to finite lattice spacing are negligible in the GMOR relation. 

\begin{figure}[t]
\begin{center}
\includegraphics[width=.49\textwidth]{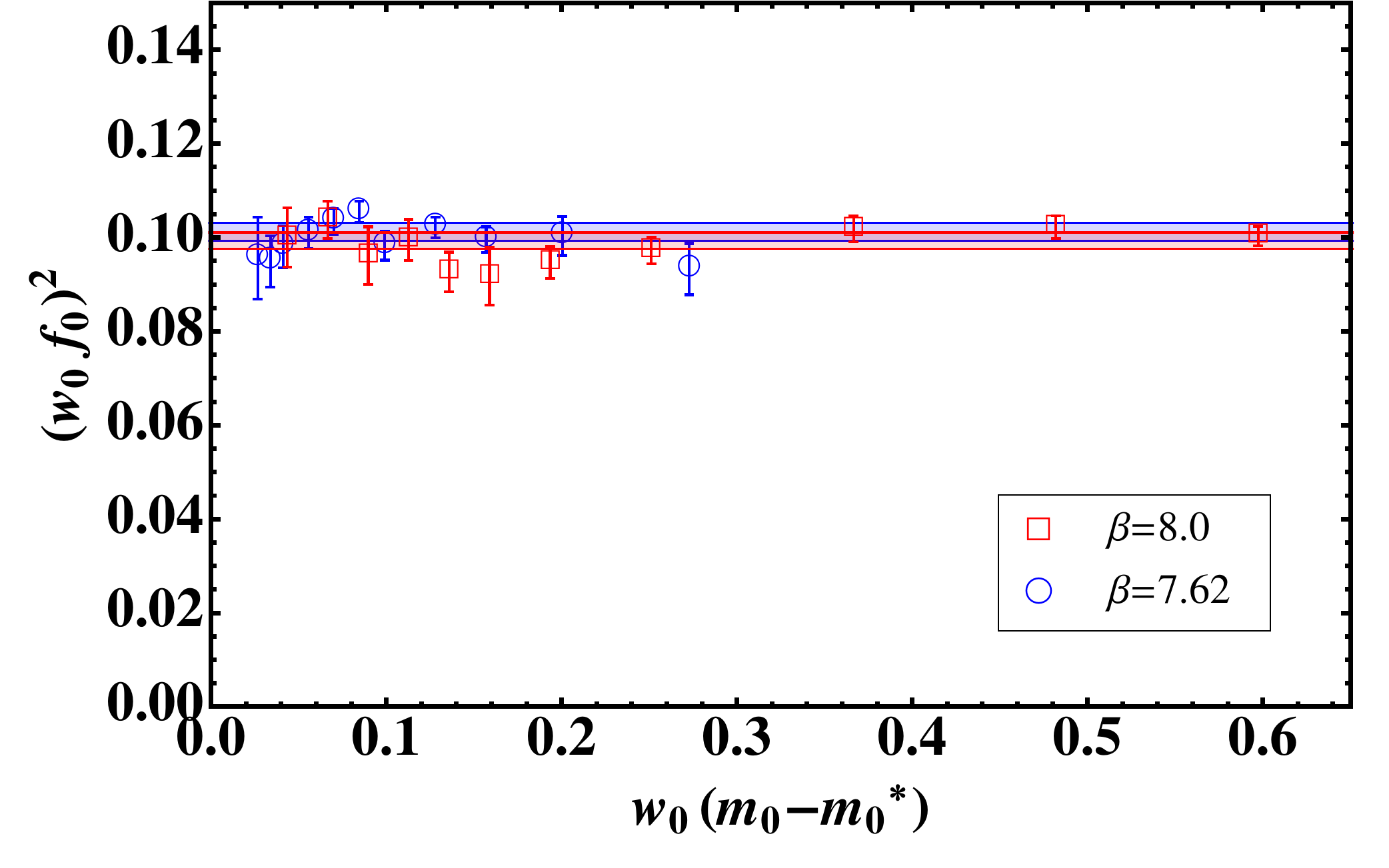}
\caption{%
Sum of the squared decay constants of the pseudoscalar, vector and axial-vector mesons. 
The coloured band denotes the (constant) fit result with statistical error 
for each lattice gauge coupling. 
}
\label{fig:squared_f0}
\end{center}
\end{figure}

Among the physical observables in the (tree-level) NLO EFT, the sum of the squared decay constants of 
pseudoscalar, vector and axial-vector mesons parameterised by $f_0^2$ is expected to be 
mass-independent. 
In principle, the tree-level result can be corrected by the chiral one-loop effects including chiral logarithms. 
This would result in a non-trivial mass dependence of $f_0^2$. 
However, this does not happen: as the NLO mass term
plays the role of the counter term that cancels the UV divergence in
the chiral logarithms, no additional mass dependence appears as long
as the tree level result is independent on the fermion mass.  
Indeed, as seen in Fig.~\ref{fig:squared_f0}, we numerically confirmed this EFT result for $f_0^2$. 
Note that the mass independence of $f_0^2$ seems to be a feature that persists even for heavy fermions. 
We therefore perform a constant fit to the data, and find
$f_0^2=0.1012(19)$ and $f_0^2 =0.0993(17)$ for $\beta=7.62$ and $8.0$, respectively. 
This result further suggests that the lattice artefacts in $f_0^2$ are negligible. 

The primary purpose of our numerical programme here is to determine the low-energy constants 
appearing in the low-energy NLO EFT developed in Sect.~\ref{sec-2}, 
and we use the quenched data to test the applicability of this approach to generic Sp(4) theories with dynamical fermions. 
We therefore perform a (uncorrelated) global fit to the data for the
masses and decay constants of pseudoscalar, vector and axial-vector
mesons using the formulae in Eq.~(\ref{eq:observables_eft}). 
The fitting ranges are restricted to the eight and six lightest fermion masses for $\beta=7.62$ and $8.0$, respectively, 
as we do not expect the NLO EFT to describe correctly the heavy fermion regime.
Using the standard fit procedure employing a $\chi^2$ minimisation, we are confronted with two technical obstacles. 
First of all, statistical uncertainties for different types of mesons
vary widely; hence, the fit results are essentially constrained by the pseudoscalar mesons. 
Secondly, the global fits are not stable, because the parameter space of the NLO EFT, 
which contains thirteen parmaters including the critical fermion mass, 
is too large for us to be able to determine the actual global minimum. 

\begin{figure}[t]
\begin{center}
\includegraphics[width=.48\textwidth]{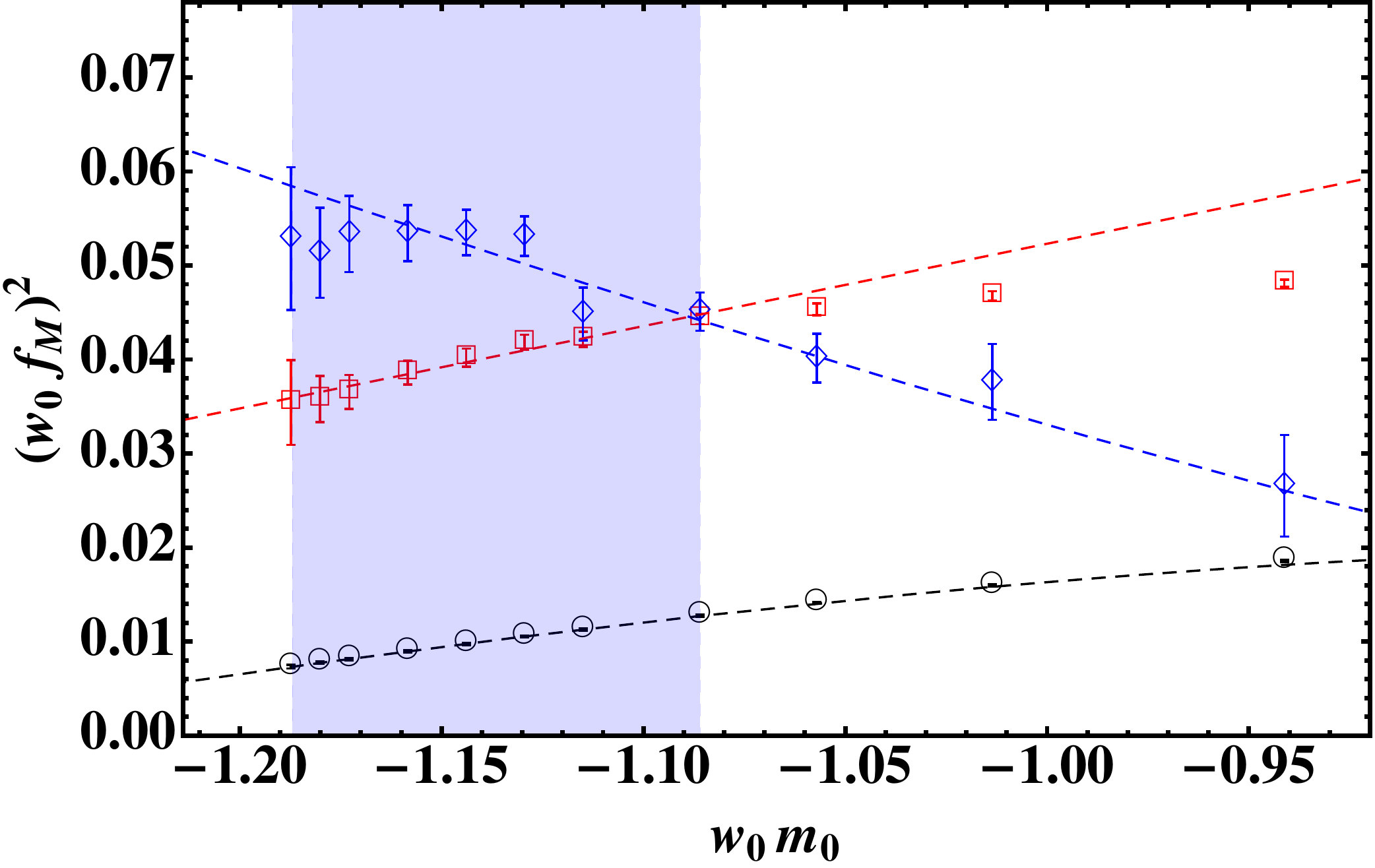}
\includegraphics[width=.48\textwidth]{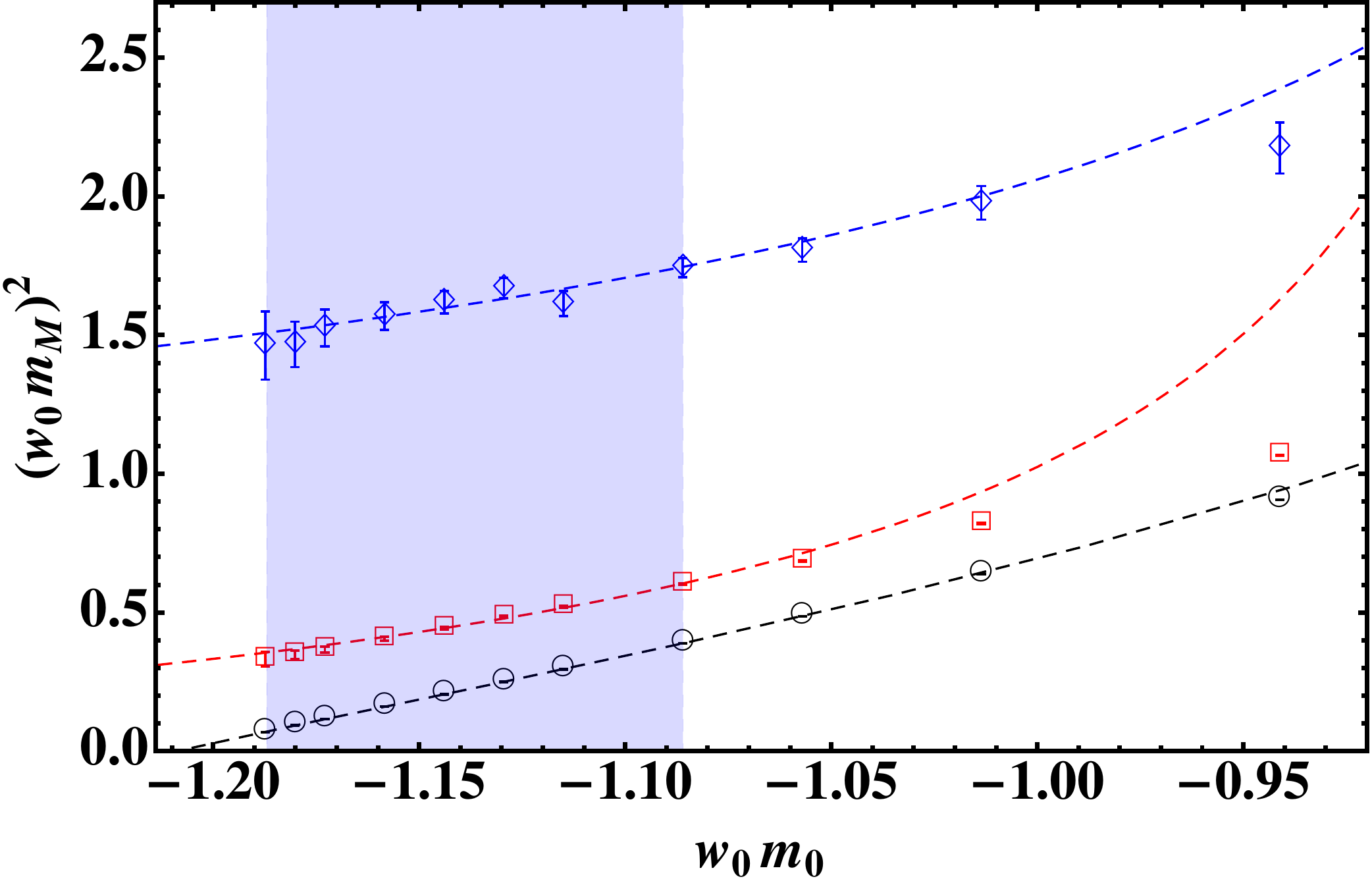}
\caption{%
Masses and decay constants of pseudoscalar (black empty circle), 
vector (red empty square) and axial-vector (blue empty diamond) mesons 
in the quenched Sp(4) at $\beta=7.62$, 
measured with respect to the fermion mass. 
Scale setting with the gradient flow method and 
perturbative one-loop matching for the decay constants are understood. 
A global fit has been performed simultaneously 
using all the measured masses and decay constants 
over the range of the bare fermion mass $-0.82\leq m_0\leq-0.75$, 
with $w_0=1.448$. The dashed lines denote the fit results.  
}
\label{fig:global_fit}%
\end{center}
\end{figure}

Despite these caveats, as the purpose of the quenched calculation is to illustrate the usage of our EFT, 
we attempt to perform the global fit to the central values. 
Because of uncontrolled systematics such as discretisation and
quenching effects, we do not expect to extract the physical
information from this explorative study. The results are illustrated
in Fig.~\ref{fig:global_fit} taking the coarse lattice ($\beta=7.62$) as
an example. The extracted fit values are 
\beq
&\kappa=-0.76,~y_3=-0.61,~y_4=3.21,~g_c=2.14,~b=-0.42,~c=-0.031,~f_1=0.74,~f_2=0.55,&\nn \\
&~v_1=0.088,~v_2=-0.31,~v=0.27,~v_5=0.40,~m_0^*=-1.21,~\chi^2/{\textrm{d.o.f}}=0.74,&
\eeq
for $\beta=7.62$ and 
\beq
&\kappa=-0.87,~y_3=-0.35,~y_4=2.75,~g_c=1.74,~b=-0.30,~c=-0.012,~f_1=0.97,~f_2=0.60,&\nn \\
&~v_1=0.034,~v_2=-0.30,~v=0.28,~v_5=0.48,~m_0^*=-1.64,~\chi^2/{\textrm{d.o.f}}=0.97,&
\eeq
for $\beta=8.0$, respectively, where all of these fit results satisfy the unitary conditions. 

The NLO EFT also allows us to compute the Peskin-Takeuchi $S$-parameter and the $\rho$-$\pi$-$\pi$ coupling 
in the massless limit,
\beq
&S/4\pi\sim0.065,~~~g_{\rho\pi\pi}^2/(48\,\pi)\sim0.9&~~\textrm{for $\beta=7.62$}, \nn \\
&S/4\pi\sim0.053,~~~g_{\rho\pi\pi}^2/(48\,\pi)\sim1.2&~~\textrm{for $\beta=8.0$}.
\eeq
The resulting $S$ parameters of Sp(4) theories are larger than 
the QCD values of $S/4\pi\sim0.025$ computed in the same way (i.e.~the zeroth sum rule, 
$S/4\pi=f_{\textrm{V}}^2/M_{\textrm{V}}^2-f_{\textrm{AV}}^2/M_{\textrm{AV}}^2$). 
This result agrees with our naive expectation based on the fact that 
the dimension of the gauge group is larger than that of QCD. 
In addition, the $S$ parameter seems to decrease as $a\rightarrow 0$. 
It is worth stressing once again that one cannot take these results
too literally because the lattice artefacts and quenching effects  
are not fully accounted for. Similarly, the large values of
$g_{\rho\pi\pi}$ are presumably due to the quenching effects.  

\begin{figure}[t]
\begin{center}
\includegraphics[width=.48\textwidth]{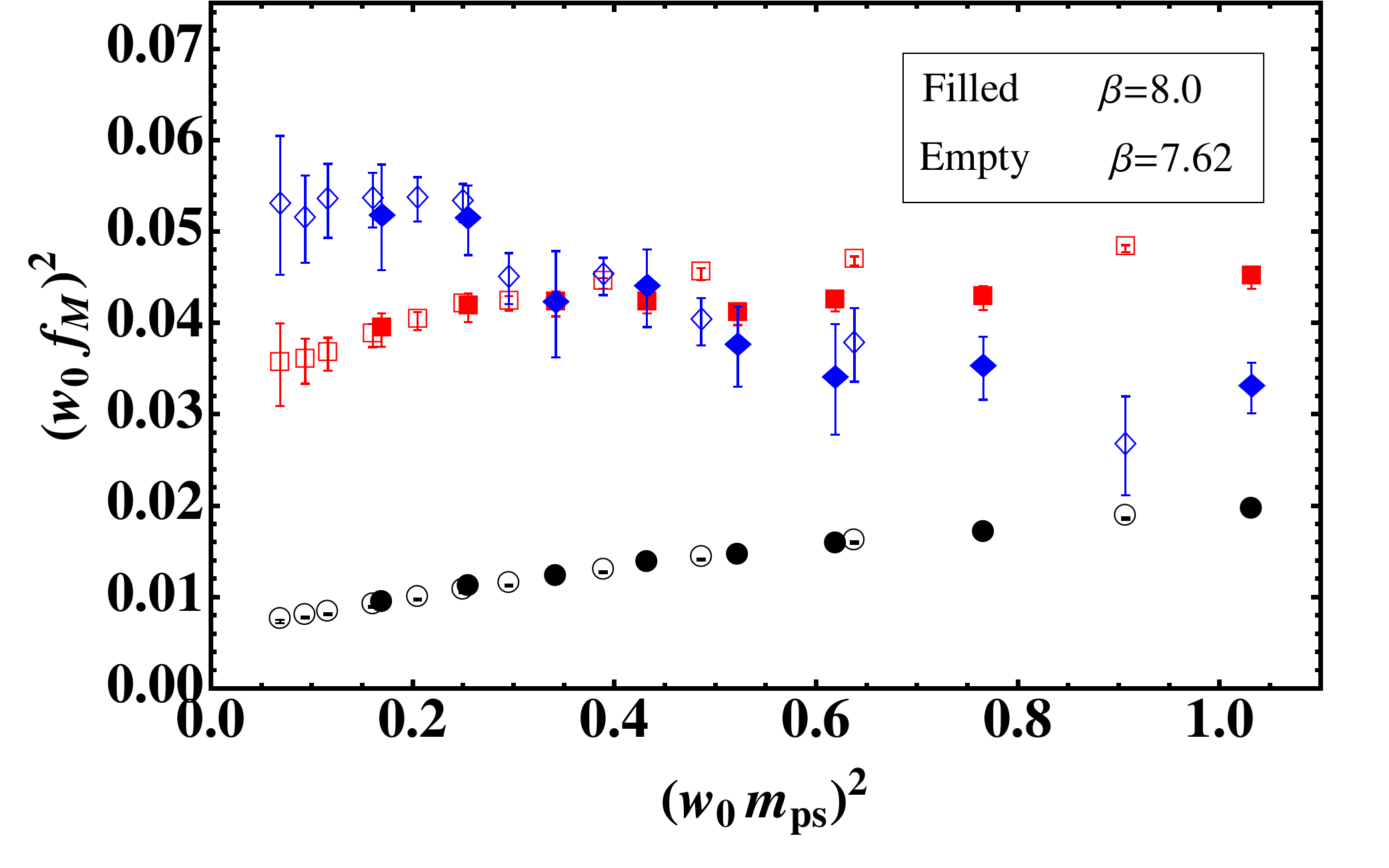}
\includegraphics[width=.48\textwidth]{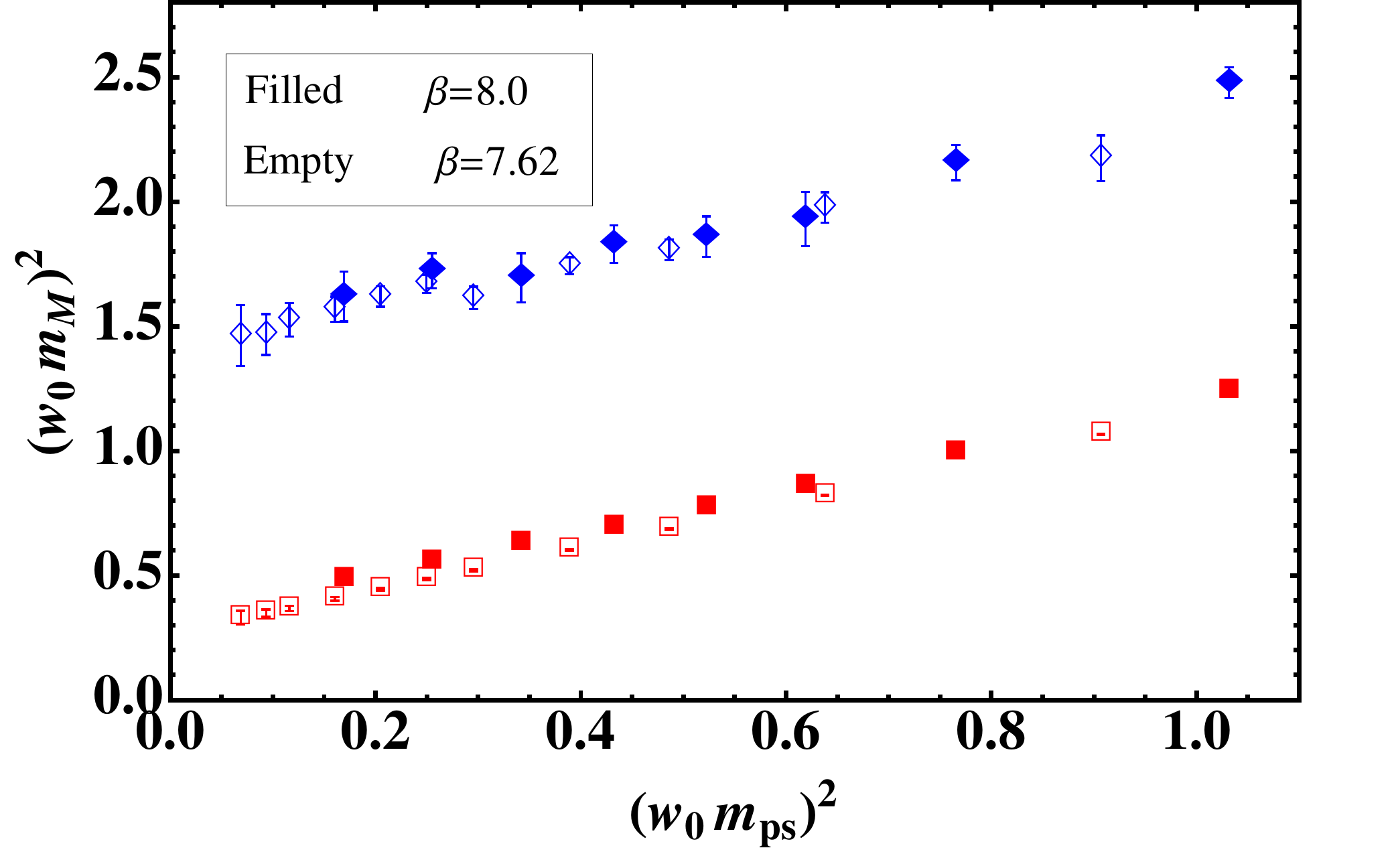}
\caption{%
Masses and decay constants of pseudoscalar (black circle), 
vector (red square) and axial-vector (blue diamond) mesons 
in the quenched Sp(4) at $\beta=7.62$ (empty) and $8.0$ (filled), 
measured with respect to the pseudoscalar mass squared. 
Scale setting with the gradient flow method and 
perturbative one-loop matching for the decay constants are understood. 
}%
\label{fig:meson_spectra}%
\end{center}
\end{figure}

As we learned from the meson spectrum in the pseudoscalar sector, 
it is useful to calculate mesonic quantities as a function of the pseudoscalar mass 
to avoid the cumbersome renormalization procedure for the fermion mass. 
As shown in Fig.~\ref{fig:meson_spectra}, 
the lattice spacing artefacts are quite large for the vector mesons, 
while those are negligible for the pseudoscalar and axial-vector mesons. 

\section{Toward dynamical simulations}
\label{dynamical}
To the best of our knowledge,
the Sp(4) theory with dynamical $N_f=2$ Wilson fermions
has not been studied before in the literature.
Therefore, first and foremost, we need to investigate the phase structure in the lattice parameter space
in order to look for potential first order bulk transitions. 
From such studies one can identify the (weak coupling) phase that is
analytically connected to the continuum theory as $a \to 0$. 
Since one cannot consider an arbitrarily small $a$ and an arbitrarily large lattice volume $V$,
the identification of the phase boundary is also an important task from the practical point of view.

\begin{figure}[t]
\begin{center}
\includegraphics[width=.79\textwidth]{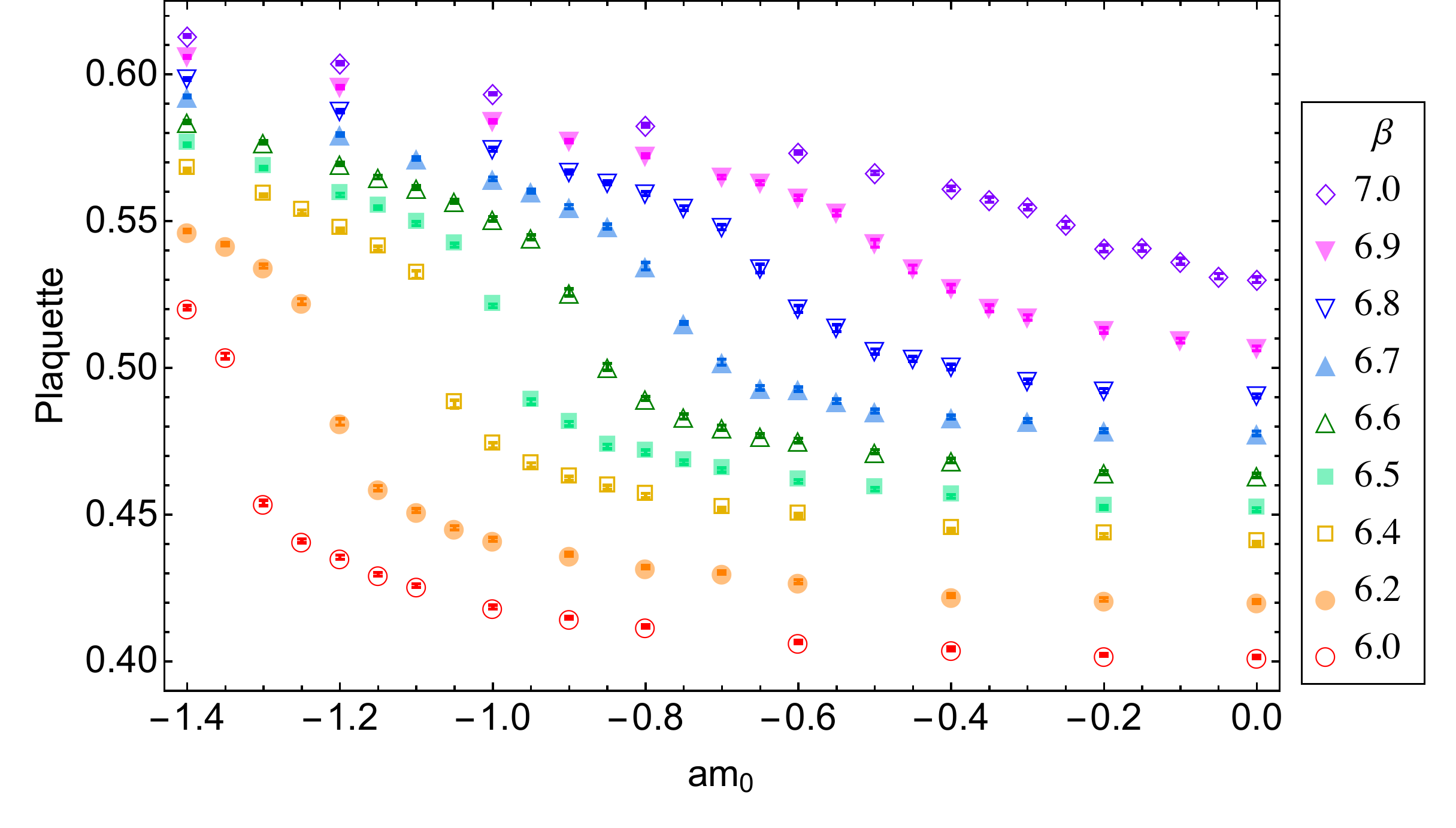}
\includegraphics[width=.48\textwidth]{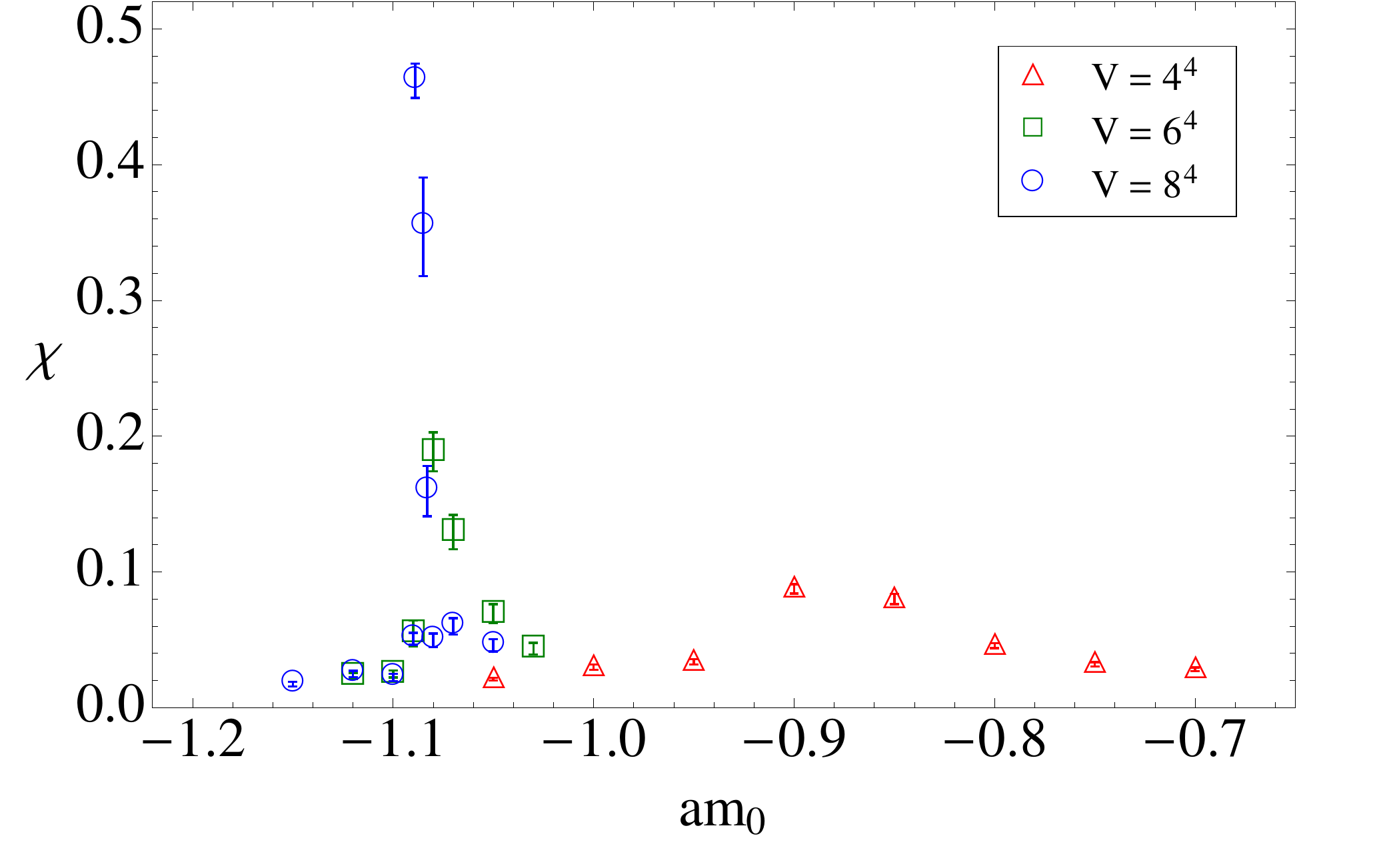}
\includegraphics[width=.48\textwidth]{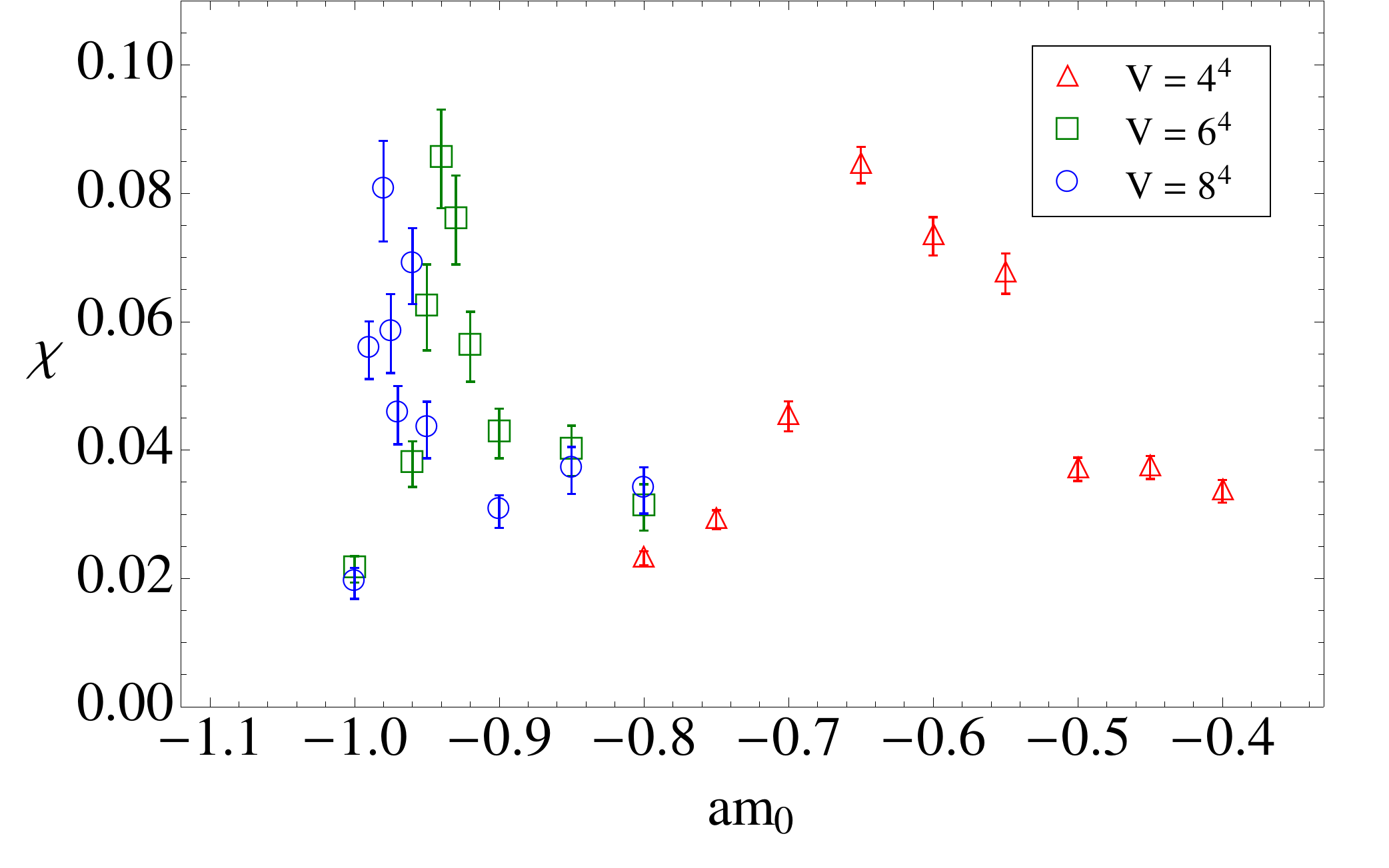}
\caption{%
(top) Structure of a phase space in the Sp(4) theory with two flavour 
(mass degenerated) dynamical Wilson fermions on a $4^4$ lattice. 
The $x$- and $y$-axes are the expectation values of the plaquette 
and the bare fermion mass. 
Different colours correspond to the lattice gauge couplings. 
(bottom) Plaquette susceptibilities versus bare fermion mass 
at $\beta=6.6$ (left) and $\beta=6.8$ (right). 
Three hypercubic lattices are considered, $4^4$ (red triangle), 
$6^4$ (green square) and $8^4$ (blue square). 
}%
\label{fig:bulkphase}%
\end{center}
\end{figure}

The order parameter associated with the lattice bulk transition is the expectation value of the plaquette.
In Fig.~\ref{fig:bulkphase}, we show the mass scan of the two-flavour Sp(4) theory using a $4^4$ lattice.
For each lattice coupling $\beta$ chosen between $6.0$ and $7.0$ we vary the fermion mass in steps of $0.1$
over the range $0.0\leq-a\,m_0\leq 1.4$.
For the region, where the average plaquette values change rapidly as $m_0$ changes,
we increase the resolution by a factor of two.
At small values of $\beta\simeq 6.6$ the plaquette values change abruptly,
indicating the existence of a bulk phase.
To accurately determine the phase boundary, we further study the finite size scaling
of the plaquette susceptibility $\chi$ at $\beta=6.6$ and $6.8$
by considering a $4^4$, $6^4$, and $8^4$ lattice. $\chi$ is defined as
\beq
\chi=\sum_x \langle \mathcal{P}(0)\,\mathcal{P}(x)\rangle
=(\langle \mathcal{P}^2\rangle-\langle \mathcal{P}\rangle^2)\,V.
\eeq
As shown in Fig.~\ref{fig:bulkphase}, the maximum values of $\chi$ for $\beta=6.6$ roughly scale with $V$
indicating a first order bulk phase transition, while those for
$\beta=6.8$ roughly constant as a function of $V$,
clearly indicating a crossover behaviour.
We therefore conclude that we can safely take the continuum limit for $\beta\geq\beta^*\simeq 6.8$.
Note that in the case of pure Sp(4) theory formulated with the standard plaquette action
bulk phase transitions do not arise.

\begin{figure}[t]
\begin{center}
\includegraphics[width=.48\textwidth]{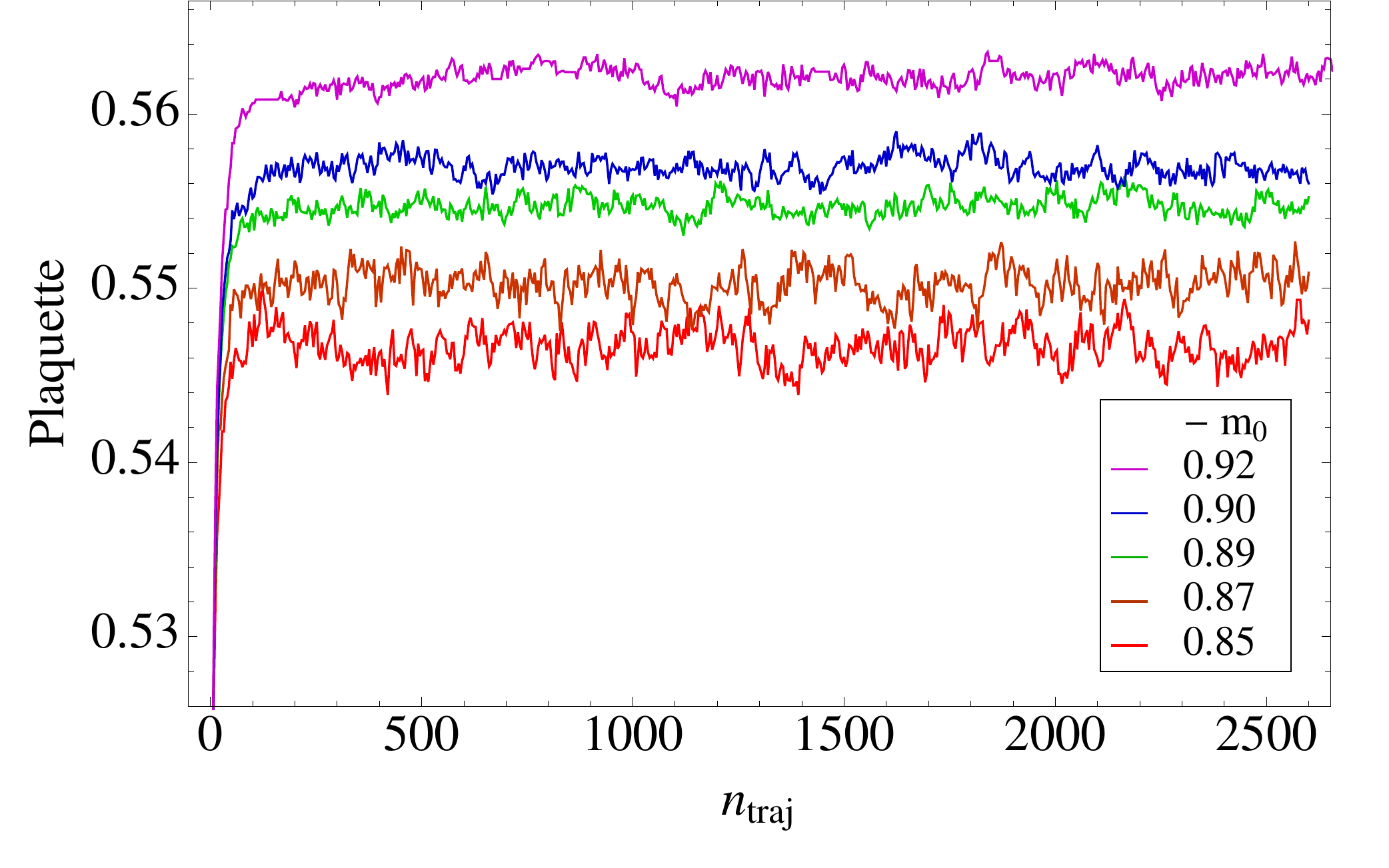}
\includegraphics[width=.48\textwidth]{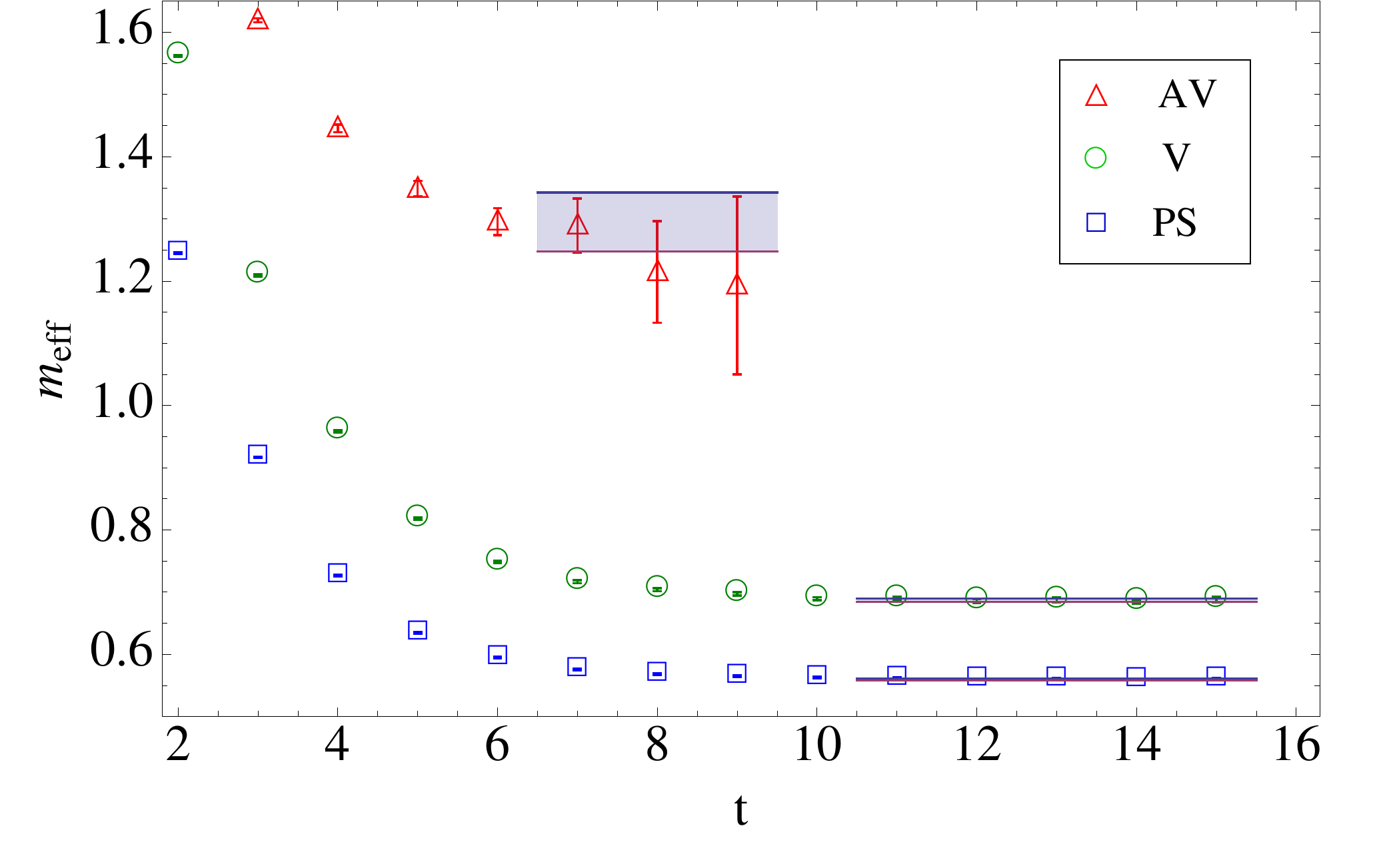}
\caption{%
(left) Plaquette trajectories at various bare fermion masses 
and lattice coupling of $\beta=6.9$. 
The lightest three fermion masses are simulated on a $32\times16^3$, 
while the remaining masses are simulated on a $24\times 12^3$. 
(right) Plot of effective masses for pseudoscalar, vector and axial-vector mesons 
at $\beta=6.9$ with $m_0=-0.9$ on a $32\times 16^3$ lattice. 
}%
\label{fig:dyn_sp4}%
\end{center}
\end{figure}

Based on our finding on the phase boundary,
we use the coarse lattice with $\beta=6.9$ to test the stability of dynamical simulations
and provide a very preliminary understandings of the meson spectrum.
We generate five ensembles using the HMC algorithm, $m_0=-0.85,\,-0.87$ on a $24\times12^3$ and
$m_0=-0.89,\,-0.9,\,-0.92$ on a $32\times16^3$ lattices.
As shown in the left panel of Fig.~\ref{fig:dyn_sp4}, the trajectories of the average plaquettes are stable and
their asymptotic values increase as the fermion mass decreases.
While the thermalisation time typically appears to be $\sim 300$,
the autocorrelation time varies between $12$ and $32$, depending on the ensembles.

To illustrate how to extract meson masses from the two-point correlation functions,
in the right panel of Fig.~\ref{fig:dyn_sp4},
we show the effective mass plots for pseudoscalar, vector and axial-vector mesons with $m_0=-0.9$.
By performing a constant fit to the plateau region at large time,
we determine the meson masses.
The masses of the pseudoscalar and vector mesons are smaller than the UV cutoff with $m_{\textrm{PS}}/m_{\textrm{V}}\sim0.8$,
while the mass of the axial-vector is at the UV scale.
Although this explorative study in the dynamical fermion case
does not provide much physical information,
in the future it will certainly guide us towards more interesting numerical findings for the $N_f=2$ Sp(4) theory,
which will enable us to perform an EFT analysis similar to the one
discussed in the previous section.

\section{Conclusion}
\label{summary}
In this contribution, we have investigated the global symmetry
breaking pattern SU(4) $\mapsto$ Sp(4) in the language of effective
field theory and, using Sp(4) gauge theory coupled to two 
fundamental Dirac fermions, we have discussed the determination of the
corresponding low-energy constants. Our investigation
provides encouraging evidence that a similar line of study can be
pursued also for the model with dynamical fermions, for which we have
presented first results in this work. As a next step in our programme,
we shall extend the lattice calculation in the dynamical system
with the longer term goal of obtaining the low-energy constants of the effective
field theory with a controlled systematics and enough statistical
accuracy to make our results relevant for phenomenology.

\section*{Acknowledgements}
We acknowledge the support of the Supercomputing Wales project, which is part-funded by the European Regional Development Fund (ERDF) via Welsh Government.

\bibliography{Lattice2017_182_LUCINI}

\end{document}